\documentclass[prd,twocolumn,showpacs,preprintnumbers,superscriptaddress,10pt,nofootinbib,floatfix]{revtex4}
\usepackage{amssymb}
\usepackage{amsmath}
\usepackage{graphicx}
\usepackage{dcolumn}
\usepackage{bm}

\input{tcilatex}

\begin{document}

\title{Pomeron loops in zero transverse dimensions}
\author{Arif~I.~Shoshi}
\email{shoshi@physik.uni-bielefeld.de}
\affiliation{Department of Physics, Columbia University, New York, NY, 10027, USA}
\affiliation{Fakult{\"a}t f{\"u}r Physik, Universit{\"a}t Bielefeld, D-33501 Bielefeld,
Germany}
\author{Bo-Wen Xiao}
\email{bowen@phys.columbia.edu}
\affiliation{Department of Physics, Columbia University, New York, NY, 10027, USA}

\begin{abstract}
We analyze a toy model which has a structure similar to that of the recently
found QCD evolution equations, but without transverse dimensions. We develop
two different but equivalent methods in order to compute the leading-order
and next-to-leading order Pomeron loop diagrams. In addition to the
leading-order result which has been derived within other toy models~\cite%
{Mueller:1994gb,Mueller:1996te}, we can also calculate the next-to-leading
order contribution which provides the $\left( \alpha _{s}^{2}\alpha Y\right)$
correction. We interpret this result and discuss its possible implications
for the four-dimensional QCD evolution.
\end{abstract}

\date{\today }
\preprint{CU-TP-1141}
\preprint{BI-TP 2005/50}
\maketitle

\section{Introduction}

There have been major breakthroughs in understanding high-density QCD
evolution recently. Several important observations have been made: (i) gluon
number fluctuations have a big effect on the evolution towards gluon
saturation~\cite{Mueller:2004se}, (ii) a connection between high-energy QCD
evolution and statistical physics models of reaction-diffusion type was
found~\cite{Iancu:2004es,Munier:2003vc} which has clarified the
interpretation of fluctuations in an event-by-event picture, and (iii) it
was realized that the relevant fluctuations are missed~\cite{Iancu:2004iy}
by the existing Balitsky-JIMWLK equations~\cite%
{Balitsky:1995ub+X,Jalilian-Marian:1997jx+X}. These observations have led to
the creation of new QCD evolution equations which include gluon number
fluctuations or Pomeron loops~\cite%
{Mueller:2005ut,Iancu:2005nj,Kovner:2005nq,Hatta:2005rn}.

The new equations describe the usual BFKL evolution, Pomeron splittings and
Pomeron mergings and, thus, through iterations, Pomeron loops. A lot is
known about the structure of the new equations~\cite%
{Kovner:2005en+X,Blaizot:2005vf+X,Marquet:2005hu,Levin:2005au,Enberg:2005cb}%
, however, because of their complexity, little about their solutions. So
far, analytical results for the energy dependence of the saturation momentum
and for the scaling behavior of the $T$-matrix at asymptotic energies~\cite%
{Mueller:2004se,Iancu:2004iy} and some numerical simulations of
approximations to the exact equations~\cite{Enberg:2005cb,Soyez:2005ha} are
available. Solutions to the new equations in the range of collider energies
are desired.

In this paper, we consider a toy model which has a structure similar to that
of the new QCD evolution equations, but which has no transverse dimensions.
This simplification allows an exact solution of the evolution equations. The
particle-particle scattering amplitude is calculated. The effects of
saturation and unitarity, or \textquotedblleft Pomeron\textquotedblright\
loops which are the characteristics of the new evolution equations, are
investigated.

A decade ago, Mueller~\cite{Mueller:1994gb} studied a similar toy model
without transverse dimensions which was motivated by the QCD dipole model~%
\cite{Mueller:1993rr+X}. He has been able to solve the toy model exactly and
calculate the $S$-matrix. Observations made by studying this toy model (e.g.
the one that fluctuations in particle numbers lead to a divergent multiple
scattering series) have been shown numerically~\cite%
{Salam:1995uy,Salam:1995zd} to be valid also in the four-dimensional QCD.
The hope that some of the results obtained in the toy model discussed in
this paper may also apply in four-dimensional QCD is one of the motivations
for this work.

In the toy model proposed by Mueller~\cite{Mueller:1994gb},
particle-particle scattering in the center-of-mass frame at total rapidity $%
Y $ was considered. In this model, each particle evolves to a dense system
through particle branching (splitting). Unitarity effects (at leading order
accuracy) are correctly described due to the multiple scatterings between
the dense systems. There are no particle mergings in the wavefunctions of
the particles in this model, therefore, the model fails to describe
saturation effects in the wavefunctions of the particles which become
important at very high rapidities $Y>Y_{s}=\frac{2}{\alpha _{s}^{2}}\ln
\frac{1}{\alpha _{s}^{2}}$. In the language of loops, only the loops
stretching over the rapidity interval greater than $Y/2$ (\textquotedblleft
large\textquotedblright\ loops) shown e.g. in Fig.~\ref{reggeon}B are
included (at leading-order accuracy and for $Y<Y_{s}$). \textquotedblleft
Large\textquotedblright\ loops are created by matching the evolution
(\textquotedblleft Pomeron\textquotedblright\ splittings) of both particle's
wavefunctions at $Y/2$. However, there are no loops inside the particle's
wavefunctions (\textquotedblleft small\textquotedblright\ loops) where the
\textquotedblleft Pomeron\textquotedblright\ splits and merges, as shown
e.g. in Fig.~\ref{reggeon}C(see, e.g. ref.\cite{Kovchegov:2005ur}), within
the rapidity interval $Y/2$. The above limitations of the toy model are, of
course, properties of the QCD dipole model~\cite{Mueller:1993rr+X} as well
and are well-known.

Mueller's toy model breaks down at $Y\simeq Y_s$. In a later work Mueller
and Salam~\cite{Mueller:1996te} did include saturation effects by imposing
boost invariance on the small-$x$ evolution in a toy model, getting an
expression for the $S$-matrix which is valid also for $Y>Y_s$. The effect of
saturation was an extension of the validity region of the result obtained
for $Y<Y_s$ in~\cite{Mueller:1994gb} also to the region where $Y>Y_s$.

Our toy model which is inspired by the new QCD evolution equations~\cite%
{Mueller:2005ut,Iancu:2005nj,Kovner:2005nq,Hatta:2005rn} allows us
to study unitarity effects as well as saturation effects in a more
refined and systematic way. We calculate the leading order (LO)
and the next-to-leading order (NLO) loop contributions to the
particle-particle scattering amplitude. The LO result, which is
the LO contribution of the \textquotedblleft
large\textquotedblright\ loops, agrees with the results
from~\cite{Mueller:1994gb,Mueller:1996te}. Our NLO result
consisting of the NLO contribution of the \textquotedblleft
large\textquotedblright\ loops and the LO contribution of the
\textquotedblleft small\textquotedblright\ loops is new. The NLO
correction with respect to the LO result is of order $\alpha
_{s}^{2}\alpha Y$. This is the main result of the paper. It tells
us that the NLO contribution is negligible as compared with the LO
contribution up to rapidities $Y\leq 1/\alpha\alpha _{s}^{2}$.

Our LO result agrees with the result derived within Mueller's toy
model up to $Y\simeq Y_s$. Beyond $Y_s$ Mueller's toy model fails
to include saturation effects. The result in~\cite{Mueller:1996te}
which involves saturation effects is the same as our LO result
until $Y\simeq 1/\alpha\alpha_s^2$ which is parametrically much
larger than $Y_s$. However, beyond $Y\simeq 1/\alpha\alpha_s^2$
the NLO corrections become important which are not taken into
account in Mueller and Salam's toy model~\cite{Mueller:1996te}.

Presumably our toy model reveals the properties of the
four-dimensional QCD evolution to some extent. In fact, its
behavior is a heuristic sign that the NLO contribution can be
neglected in the QCD evolution as well as long as $Y \ll
1/\alpha\alpha_s^2$. Such a NLO correction seems to appear also in
real QCD.
Considering the one-loop diagram for example, the LO contribution in QCD is $%
\left( \alpha _{s}^{2}\right) ^{2}\exp \left( 2\alpha Y\right) $
which comes from varying the size of the loop from $0$ to $Y$,
where $\alpha =\alpha _{P}-1=\frac{4\alpha _{s}N_{c}}{\pi }\ln 2$.
The NLO contribution $\left( \alpha _{s}^{2}\right) ^{2}\alpha
Y\exp \left( \alpha Y\right) $ comes from changing the location of
a fixed zero-size loop from $0 $ to $Y$. The suppression of the
NLO contribution by the factor $\alpha Y/e^{\alpha Y}$ directly
leads to the general correction $\alpha _{s}^{2}\alpha Y$ (the
same happens in our toy model).

The paper is organized as follows: In Sec.~\ref{Sec_The_toy_model}, we show
the toy model. Two methods are developed in order to solve the toy model and
some discussion is provided on the consequences of the NLO corrections. They
are explained in Sec.~\ref{solution_TM}. In the Appendices we show the
Lorentz-invariance of the toy model and give the solution to another
commonly used zero-dimensional model. Finally, we discuss the results and
give the conclusions.

\section{A toy model with \textquotedblleft Pomeron\textquotedblright\ loops}

\label{Sec_The_toy_model} In this section, we consider a toy model without
transverse dimensions which has a structure similar to that of the recently
found four-dimensional QCD evolution equation~\cite%
{Mueller:2005ut,Iancu:2005nj,Kovner:2005nq,Hatta:2005rn}. The toy model
describes the rapidity evolution of the particle number $n$ of a system. The
dynamics in zero-transverse dimensions is given by the following Langevin
equation\footnote{%
We provide the solution to the zero transverse dimensional sFKPP equation
with a different noise term $\sqrt{2(\alpha \widetilde{n}-\beta \widetilde{n}%
^2)}\ \nu (y)$ in Appendix.~\ref{other_Langevin_equation}}
\begin{eqnarray}
\frac{d\widetilde{n}}{dy} = \alpha \widetilde{n}-\beta \widetilde{n}^{2}+
\sqrt{2\alpha \widetilde{n}}\ \nu (y)\   \label{Langevin}
\end{eqnarray}
where $\nu (y)$ is a Gaussian white noise: $\left\langle \nu
(y)\right\rangle =0$ and $\left\langle \nu (y)\nu (y^{\prime })\right\rangle
=\delta (y-y^{\prime })$. Eq.~(\ref{Langevin}) is known as the zero
transverse dimensional \textquotedblleft Reggeon field
theory\textquotedblright\ equation and it contains the projectile-target
duality (see Appendix~\ref{duality}). The structure of Eq.~(\ref{Langevin})
is obvious: the first term on the r.h.s. represents the growing of the
particles, the second term describes particle recombinations which limit the
growth at a maximum occupancy, and the third term describes the particle
number fluctuations. These are essentially the main ingredients of the real
QCD equations.

With Ito's calculus, one can write Eq.~(\ref{Langevin}) into an infinite
hierarchy of coupled evolution equations,
\begin{eqnarray}
\frac{dn^{(k)}}{dy}=k\alpha n^{(k)}+k(k-1)\alpha n^{(k-1)}-k\beta n^{(k+1)}
\ ,  \label{h1}
\end{eqnarray}
where $n^{(k)}=\left\langle \colon \widetilde{n}^{k}\colon \right\rangle $
is the expectation value of $k$-particles during the evolution and it should
be understood as a normal ordered number operator (i.e. the factorial moment
$\left\langle \widetilde{n}(\widetilde{n}-1)\cdots (\widetilde{n}%
-k+1)\right\rangle $) according to its physical interpretation. Here, $%
k\alpha n^{(k)}$ is the growth term, $k(k-1)\alpha n^{(k-1)}$ the
fluctuation term, and $k\beta n^{(k+1)}$ the recombination term. Eqs.~(\ref%
{Langevin}, \ref{h1}) are the zero-transverse-dimensional analog of the real
QCD equations (see~\cite{Iancu:2004iy,Iancu:2005nj}). A simple derivation
and a discussion in the context of QCD evolution of eqs.~(\ref{Langevin}, %
\ref{h1}) can be found in~\cite{Iancu:2004iy}. The solution to another
widely discussed Langevin equation is shown in Appendix~\ref%
{other_Langevin_equation}.

\section{Solution to the toy model}

\label{solution_TM} Eq.~(\ref{h1}) is quite complicated since $n^{(k)}$ is
coupled with $n^{(k-1)}$ and $n^{(k+1)}$ in the evolution, i.e., Eq.~(\ref%
{h1}) is just a particular equation within an infinity hierarchy of coupled
equations. However, the following observation helps to solve Eq.~(\ref{h1}):
one can neglect the recombination term if one starts with a dilute object at
the beginning of rapidity evolution, on the other hand, one can see that $%
n^{(k)} = (\alpha/\beta)^{(k)}$ is an approximate fixed point of Eq.~(\ref%
{h1}) in extremely high rapidity and $\frac{\beta }{\alpha
}=\alpha _{s}^{2}<<1$ limit. This will be more clear if one thinks
about this in terms of the evolution equation of $T^{(k)}$(one can
define $T^{(k)}=(\alpha
_{s}^{2})^{k}n^{(k)}$.): $\frac{dT^{(k)}}{dy}=\alpha %
\left[ kT^{(k)}-kT^{(k+1)}+k(k-1)\alpha _{s}^{2}T^{(k-1)}\right] \ , $ in
which it is easy to find $\lim_{Y\rightarrow \infty }T^{(k)}=1$. Therefore,
if one starts with a dilute object at $Y=0$ ($n\ll n_s=1/\alpha _{s}^{2}$),
then one can consider the evolution equation without the recombination term
at the early stages of the evolution, and turn on the recombination term as
a perturbation at the end of the evolution which forms the \textquotedblleft
Pomeron\textquotedblright\ loops. Following this philosophy, we have found
two different methods to solve Eq.~(\ref{h1}) which we will present in the
following subsections. Before entering the calculations, we should state
that we always work in a parametrical limit in which $\alpha^{2}_{s}%
\rightarrow0$ and $\alpha Y \rightarrow \infty$, but $\alpha^{2}_{s}e^{%
\alpha Y}$ is fixed and finite.

\subsection{Method I: Integral Representation}

\label{integral_representation} Eq.~(\ref{h1}) without the recombination
term,
\begin{eqnarray}
\frac{dn_{0}^{(k)}}{dy}=k\alpha n_{0}^{(k)}+k(k-1)\alpha n_{0}^{(k-1)} \ ,
\label{h2}
\end{eqnarray}
can be solved exactly if given some physically chosen initial conditions. In
this paper, we consider the evolution of a single particle for which the
initial conditions are $\left. n_{0}^{(1)}\right\vert _{y=0}=N$ and $\left.
n_{0}^{(k)}\right\vert _{y=0}^{k>1}=0$. The solution to Eq.~(\ref{h2}) with
the above initial conditions reads
\begin{eqnarray}
n_{0}^{(k)}(y)=Nk!e^{k\alpha y}(1-e^{-\alpha y})^{k-1}  \label{split}
\end{eqnarray}%
and corresponds to the $k$-particle density obtained via splitting from a
single particle after the evolution over the rapidity $y$.
\begin{figure}[tbp]
\begin{center}
\includegraphics[width=\columnwidth]{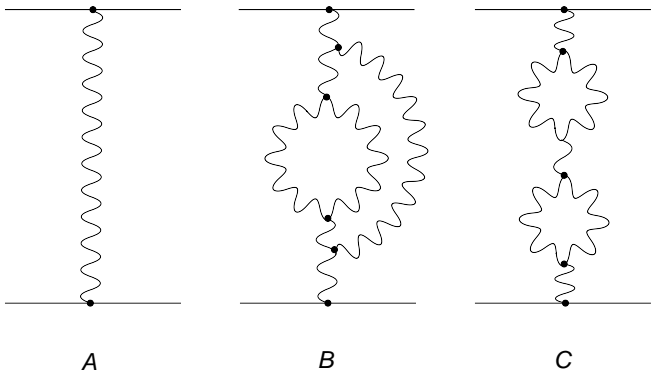}
\end{center}
\caption[*]{Three kinds of reggeon graphs, in which the curly lines
represent BFKL ladders. Diagram (A) is a simple BFKL ladder exchanged
between the two particles (horizontal lines); Diagram (B) is the LO Pomeron
loop diagram which has been calculated in ref.~\protect\cite%
{Mueller:1994gb,Mueller:1996te,Salam:1995zd}; Diagram (C) is the NLO pomeron
loop diagram which was not included in the dipolar approach of ref.~%
\protect\cite{Mueller:1994gb,Mueller:1996te,Salam:1995zd} while it can be
computed in this toy model.}
\label{reggeon}
\end{figure}
To turn on the perturbation, let us start with the first equation in the
infinite hierarchy~(\ref{h1}),
\begin{eqnarray}
\frac{dn_{1}^{(1)}}{dy}=\alpha n_{1}^{(1)}-\beta n_{0}^{(2)} \ ,  \label{h3}
\end{eqnarray}
which has the solution
\begin{eqnarray}
n_{1}^{(1)}(Y) &=&n_{0}^{(1)}(Y)-\beta \int_{0}^{Y}dye^{\alpha
(Y-y)}n_{0}^{(2)}(y)  \notag \\
&=&Ne^{\alpha Y}-2\frac{\beta }{\alpha }Ne^{2\alpha Y}(1-\frac{\alpha Y}{%
e^{\alpha Y}}-\frac{1}{e^{\alpha Y}}).  \label{s}
\end{eqnarray}
It is straight forward to recognize that the first term in Eq.~(\ref{s}) is
the usual BFKL term, and the second term corresponds to the one-loop diagram
with the right intrinsic minus sign. One can also see that $\frac{\alpha Y}{%
e^{\alpha Y}}$ is the NLO contribution in the parametrical limit we
mentioned above.
\begin{widetext}
To get the LO two-loop diagram shown in Fig.~\ref{reggeon}~B, we start with
the second equation in Eq.~(\ref{h1}),
\begin{eqnarray}
\frac{dn_{1}^{(2)}}{dy}=2\alpha n_{1}^{(2)}+2\alpha n_{0}^{(1)}-2\beta
n_{0}^{(3)},  \label{s0}
\end{eqnarray}
whose solution is
\begin{eqnarray}
n_{1}^{(2)}(Y) &=& n_{0}^{(2)}(Y)- 2 \beta \int_{0}^{Y}dye^{2
\alpha (Y-y)}\ n_{0}^{(3)}(y)  \notag \\
&=&2 N\,e^{2 \alpha Y}(1-e^{-\alpha Y}) - 2 \cdot 3! \frac{\beta}{\alpha} N
e^{3 \alpha Y}(1-\frac{2 \alpha Y} {e^{\alpha Y}} + ... ) \ ,  \label{s1}
\end{eqnarray}
then turn on the perturbation, $\frac{dn_{2}^{(1)}}{dy}=\alpha
n_{2}^{(1)}-\beta n_{1}^{(2)}$, in order to take into account the
recombination from $n^{(2)}$ to $n^{(1)}$ (in addition to the
recombination from $n_{0}^{(3)}$ to $n^{(2)}$ given by
Eq.~(\ref{s0})), which leads to
\begin{eqnarray}
n_{2}^{(1)}(Y) &\!=\!& n_{0}^{(1)}(Y)- \beta
\int_{0}^{Y}dye^{\alpha (Y-y)}\ n_{1}^{(2)}(y)  \notag \\
&\!=\!& N\,e^{\alpha Y} - 2 \frac{\beta}{\alpha} N\,e^{2 \alpha Y}(1-\frac{%
\alpha Y}{e^{\alpha Y}} - \frac{1}{e^{\alpha Y}}) + 3! \left(\frac{\beta}{%
\alpha}\right)^2 N e^{3 \alpha Y}(1-\frac{4 \alpha Y}{e^{\alpha Y}} + {{%
\mathcal{O}}}(\frac{1}{e^{\alpha Y}}) ) \ .  \label{ss2}
\end{eqnarray}
The third term in Eq.~(\ref{ss2}) represents the LO two-loop
diagram shown in Fig.~\ref{reggeon}~B.

In order to calculate the NLO two-loop diagram shown in
Fig.~\ref{reggeon}~C, one has to enforce a recombination to happen
after a splitting. Thus, starting with $n_0^{(1)}=N e^{\alpha Y}$,
one splitting given by $n_{2}^{(0)}(Y) = 2 \alpha
\int_{0}^{Y}dye^{2 \alpha (Y-y)}\ n_{0}^{(1)}(y)$, followed by one
merging $\Delta n_1^{(1)} = -\beta \int_{0}^{Y}dye^{\alpha (Y-y)}\
n_{0}^{(2)}(y)$, yields the result for the one-loop diagram. A
further splitting and a subsequent merging gives the result for
the NLO two-loop diagram,
\begin{eqnarray}
\Delta n^{(1)}(Y)= 3!\left( \frac{\beta }{\alpha }\right)^{2} Ne^{3\alpha
Y}\left( \frac{2}{3} \frac{\alpha Y}{e^{\alpha Y}} + \mathcal{O}\left( \frac{%
1}{e^{\alpha Y}}\right) \right) \ .
\end{eqnarray}

In fact, in order to calculate any loop diagram, one just needs to
know the merging and splitting vertices which according to
Eq.~(\ref{h1}) read, respectively,
\begin{eqnarray}
\Delta n^{(k)}_{m}(Y) &=&-k\beta \int_{0}^{Y}dye^{k\alpha
(Y-y)}\Delta
n^{(k+1)}(y) \ ,  \label{merging} \\
\Delta n^{(k)}_{s}(Y) &=&k\left( k-1\right) \alpha
\int_{0}^{Y}dye^{k\alpha (Y-y)}\Delta n^{(k-1)}(y) \ .
\label{splitting}
\end{eqnarray}

For particle-particle scattering, the $\left(k+1\right)$-th order
diagram contains $k$-splittings and $k$-mergings: The
leading-order diagram corresponds to the diagram in which all
$k$-splittings occur before all the $k$-mergings, the
next-to-leading order diagram is always the diagram in which
exactly one merging takes place before one splitting. The sum over
all LO and NLO loop diagrams gives the LO and NLO contributions to
the $T$-matrix, respectively.

For particle-particle scattering, the sum over the LO loop
diagrams (leading and next-to-leading contribution) reads
\begin{eqnarray}
n_{LO}^{(1)}(Y) = \dsum_{k=1}^{\infty }\left(-1\right)^{k+1}
N\cdot k!e^{k\alpha Y} \left(\frac{\beta }{\alpha }\right)^{k-1}
\left(1 -
(k-1)^2 \ \frac{\alpha Y}{e^{\alpha Y}} + \mathcal{O}\left( \frac{1}{%
e^{\alpha Y}}\right) +\mathcal{O}\left( \left( \frac{\alpha Y}{e^{\alpha Y}}%
\right)^{2}\right) \right) \ ,
\label{LO_Loop}
\end{eqnarray}
while the sum over the NLO loop diagrams (leading contribution) is
\begin{eqnarray}
n_{NLO}^{(1)}(Y) &=& \dsum_{k=1}^{\infty }\left(-1\right)^{k+1}
N\cdot k!e^{k\alpha Y} \left(\frac{\beta }{\alpha }\right)^{k-1}
\left(
\frac{(k\!-\!1)(k\!-\!2)^2}{k} \frac{\alpha Y}{e^{\alpha Y}} + \mathcal{O}%
\left( \frac{1}{e^{\alpha Y}}\right) + \mathcal{O}\left(\!\left( \frac{%
\alpha Y}{e^{\alpha Y}}\right)^{2}\right) \!\right) .
\label{NLOloop}
\end{eqnarray}
The resulting $S$-matrix, $S \equiv 1-T$ (with $T = \alpha_s^2
(n^{(1)}_{LO}+n^{(1)}_{NLO})$), which includes LO and NLO loop
diagrams, can be written in the form
\begin{eqnarray}
S(Y) = \dsum_{k=0}^{\infty }\left( -1\right) ^{k}\cdot k!\left(
\alpha _{s}^{2}e^{\alpha Y}\right)^{k} \Big\{1 +\alpha
_{s}^{2}\alpha Y\cdot \left( 3k^{2}-k\right)+ \alpha _{s}^{2}\cdot
f_{1}(k)+\left( \alpha _{s}^{2}\alpha Y\right) ^{2}\cdot
f_{2}(k)+...\Big\} \ , \label{smatrix}
\end{eqnarray}
where we have put $\frac{\beta}{\alpha}= \alpha_s^2$ in close
analogy with the four-dimensional QCD and $N=1$. $f_{1}(k)$ and $
f_{2}(k)$ are functions of $k$, which are too complicated to be
calculated in this method. The series in Eq.~(\ref{smatrix}) is a
divergent series, however, it is Borel-summable. After the
Borel-summation, the expression for the $S$-matrix becomes rather
complicated,
\begin{eqnarray}
S(Y) &=& \frac{1}{\alpha _{s}^{2}e^{\alpha Y}}\exp \left(
\frac{1}{\alpha _{s}^{2}e^{\alpha Y}}\right) \Gamma \left(
0,\frac{1}{\alpha _{s}^{2}e^{\alpha Y}}\right)\nonumber\\
&+& (\alpha^2 _s \alpha Y) \left\{ \frac{1}{\alpha_{s}^{2}e^{\alpha Y}}%
\exp \left( \frac{1}{\alpha _{s}^{2}e^{\alpha Y}}\right) \Gamma \left( 0,%
\frac{1}{\alpha _{s}^{2}e^{\alpha Y}}\right) \left[ 4+
\frac{10}{\alpha_s^2 e^{\alpha Y}} + \frac{3}{(\alpha_s^2
e^{\alpha Y})^2} \right] -\left[ \frac{7}{\alpha_{s}^{2}e^{\alpha
Y}} + \frac{3}{(\alpha_{s}^{2}e^{\alpha Y})^2}\right] \right\}
\nonumber\\&+& (\alpha_s^2)
\left\{\tilde{f_1}(\alpha_{s}^{2}e^{\alpha Y})\right\} +
(\alpha_{s}^{2}\alpha Y)^2
\left\{\tilde{f_2}(\alpha_{s}^{2}e^{\alpha Y})\right\} + ... \ ,
\label{S_exact}
\end{eqnarray}
however, it gets considerably simplified when $\alpha
_{s}^{2}e^{\alpha Y}$ becomes large,
\begin{eqnarray}
S(Y)=\frac{1}{\alpha _{s}^{2}e^{\alpha Y}}\exp \left(
\frac{1}{\alpha _{s}^{2}e^{\alpha Y}}\right) \Gamma\left(
0,\frac{1}{\alpha
_{s}^{2}e^{\alpha Y}}\right) \left[ 1+4\alpha _{s}^{2}\alpha Y+\mathcal{O}%
\left( \frac{1}{\alpha _{s}^{2}e^{\alpha Y}}\right) \right] \ ,
\label{approx}
\end{eqnarray}
where $\Gamma(0,x)$ is the incomplete gamma function. The NLO
correction term $4\alpha _{s}^{2}\alpha Y$ inside the brackets
with respect to the leading contribution is new.
\end{widetext}
%----------------------------------------------------------------------

\subsection{Method II: $\protect\omega$-representation}

In this section, we develop another more powerful method, the $\omega$%
-representation, to solve the toy model. This method exhibits the
pole-structure of the toy model which bears some resemblance to the
pole-structure of the BFKL equation in QCD. The $\omega $-representation is
equivalent to the integral representation method in Sec.~\ref%
{integral_representation}, however, it enables us to go one step further by
resumming all the $\alpha _{s}^{2}\alpha Y$ terms in the $S$-matrix.
Therefore, this method allows us to get an expression for the $S$-matrix
which is assumed to be valid up to the rapidity $Y_{c2}=\frac{1}{\alpha
_{s}^{4}\alpha }$ where the next-to-next-to-leading order $\left( \mathcal{O}%
\left( \alpha _{s}^{4}\right) \alpha Y\right) $ becomes of order one.

Let us show how to solve Eq.~(\ref{h1}) with the $\omega$-representation
method. One starts with the Laplace transform of Eq.~(\ref{h1})
\begin{widetext}
\begin{eqnarray}
\omega n^{(k)}(\omega )-\left. n^{(k)}(t)\right\vert _{t=0}=kn^{(k)}(\omega
)+k(k-1)n^{(k-1)}(\omega )-k\alpha _{s}^{2}n^{(k+1)}(\omega ) \ ,  \label{lp}
\end{eqnarray}
where we have used the definitions $t=\alpha y$ and
$n^{(k)}(\omega )=\int_{0}^{\infty}dtn^{(k)}(t)e^{-\omega t}$.
With the initial condition $\left. n_{0}^{(1)}\right\vert
_{y=0}=N$ and $\left. n_{0}^{(k)}\right\vert _{y=0}^{k>1}=0$,
Eq.~(\ref{lp}) can be written into a matrix form:
\begin{eqnarray}
\left(
\begin{array}{ccccc}
\omega -1 & \alpha _{s}^{2} & 0 & 0 & 0 \\
-2 & \omega -2 & 2\alpha _{s}^{2} & 0 & 0 \\
0 & ... & ... & ... & 0 \\
0 & 0 & -k(k-1) & \omega -k & k\alpha _{s}^{2} \\
0 & 0 & 0 & ... & ...%
\end{array}%
\right) \cdot \left(
\begin{array}{c}
n^{(1)}(\omega ) \\
n^{(2)}(\omega ) \\
... \\
n^{(k)}(\omega ) \\
...%
\end{array}%
\right) =\left(
\begin{array}{c}
N \\
0 \\
0 \\
0 \\
0%
\end{array}%
\right) \ .  \label{matrix}
\end{eqnarray}
It is fairly complicated to solve this equation directly because
of the infinite dimension it has. Nevertheless, there are two ways
to attack this problem. First, we can follow the idea that the
$\alpha _{s}^{2}$ is small, which reduces the importance of
recombination at the beginning of the evolution. Then we can solve
the above linear algebra equation perturbatively which will prove
that this approach is completely equivalent to the integral
representation method we developed in the previous section.
Second, noticing that Eq.~(\ref{matrix}) has an exact solution at
finite dimensions, we can truncate the Eq.~(\ref{matrix}) to a
$k\times k$ linear algebra equation and then take the
$k\rightarrow \infty $ limit to get the solution of our infinite
dimension matrix.

First, let us turn off the perturbation and only solve the unperturbed
matrix to get the corresponding $n_{0}^{(k)}(\omega )$. One can easily find
the solution to be $n_{0}^{(k)}(\omega )=\frac{N(k-1)!k!}{\left( \omega
-1\right) \cdot \left( \omega -2\right) \cdot \cdot \cdot \left( \omega
-k\right) }=\frac{N(k-1)!k!\Gamma \left( \omega -k\right) }{\Gamma \left(
\omega \right) }$. After the inverse Laplace transformation,
\begin{eqnarray}
n_{0}^{(k)}(\alpha y) &=&\frac{1}{2\pi i}\int_{s-i\infty
}^{s+i\infty }d\omega e^{\omega \alpha y}n_{0}^{(k)}(\omega ), \\
&=&Nk!e^{k\alpha y}(1-e^{-\alpha y})^{k-1} \ ,
\end{eqnarray}%
the result is exactly the same as what we have obtained in the previous
section.

Turning on the perturbation, we find that the first-order
correction $\Delta n_{1}^{(k)}(\omega )$ to $n_{0}^{(k)}(\omega )$
satisfies the equation:
\begin{eqnarray}
\left(
\begin{array}{ccccc}
\omega -1 & 0 & 0 & 0 & 0 \\
-2 & \omega -2 & 0 & 0 & 0 \\
0 & ... & ... & 0 & 0 \\
0 & 0 & -k(k-1) & \omega -k & 0 \\
0 & 0 & 0 & ... & ...%
\end{array}%
\right) \cdot \left(
\begin{array}{c}
\Delta n_{1}^{(1)}(\omega ) \\
\Delta n_{1}^{(2)}(\omega ) \\
... \\
\Delta n_{1}^{(k)}(\omega ) \\
...%
\end{array}%
\right) =\left(
\begin{array}{c}
-\alpha _{s}^{2}n_{0}^{(2)}(\omega ) \\
-2\alpha _{s}^{2}n_{0}^{(3)}(\omega ) \\
... \\
-k\alpha _{s}^{2}n_{0}^{(k+1)}(\omega ) \\
...%
\end{array}%
\right) \ .
\end{eqnarray}
\end{widetext} The general solution of this equation is too complicated to
be written down. However, the strategy of how to get it is straightforward.
For example, we can easily obtain $\Delta n_{1}^{(1)}(\omega )=\frac{%
-2\alpha _{s}^{2}N}{\left( \omega -1\right) ^{2}\cdot \left( \omega
-2\right) }$ which corresponds to the one-loop diagram. It yields the same
result as given in Eq.~(\ref{s}) after the inverse Laplace transformation.
This example shows that the result for each diagram calculated in the
previous section becomes pretty much simplified in the $\omega$%
-representation. To see this more clearly, one can do the Laplace transform
of Eq.~(\ref{merging}) and Eq.~(\ref{splitting}) which yields
\begin{eqnarray}
\Delta n^{(k)}_{m}(\omega ) &=&\frac{-k\alpha _{s}^{2}}{\omega -k}\Delta
n^{(k+1)}(\omega ) \ , \\
\Delta n^{(k)}_{s}(\omega ) &=&\frac{k\left( k-1\right) }{\omega -k}\Delta
n^{(k-1)}(\omega ) \ .
\end{eqnarray}
According to these equations, one can directly write down the contribution
for any diagram in this $\omega$-representation since the merging of $k+1$%
-particles to $k$-particles is nothing but adding a $\frac{1}{\omega -k}$
pole with a prefactor $-k\alpha _{s}^{2}$ in $\omega$-complex plane and the
splitting of $k-1$-particles to $k$-particles is nothing but adding a $\frac{%
1}{\omega -k}$ pole with a prefactor $k\left( k-1\right)$. Although the
problem has been simplified a lot in the $\omega$-representation, the exact
solution to Eq.~(\ref{matrix}) is nevertheless unavailable this way.

\begin{widetext}
We get the solution to Eq.~(\ref{matrix}) as follows: We first truncate the
infinite dimension matrix~(\ref{matrix}) into a finite $k\times k$ matrix,
then exactly solve the amputated equation. Let us show the solutions for the
first four $k$-values:

\begin{itemize}
\item $k=1$ case: One obtains $n_{1\times 1}^{(1)}(\omega )=\frac{N}{\omega
-1}$ which transforms to $n_{0}^{(1)}(\alpha Y)=Ne^{\alpha Y}$.

\item $k=2$ case: One obtains $n_{2\times 2}^{(1)}(\omega )=\frac{N\left(
\omega -2\right) }{\left( \omega -1\right) \left( \omega -2\right)
+2\alpha _{s}^{2}}$ which transforms to
\begin{eqnarray}
n_{1}^{(1)}(\alpha Y)=\frac{1-2\alpha _{s}^{2}}{1-4\alpha _{s}^{2}}%
Ne^{\left( 1+2\alpha _{s}^{2}\right) \alpha Y}-\frac{2\alpha _{s}^{2}}{%
1-4\alpha _{s}^{2}}Ne^{\left( 2-2\alpha _{s}^{2}\right) \alpha Y}.
\end{eqnarray}
\item $k=3$ case: One obtains $n_{3\times 3}^{(1)}(\omega )=\frac{N\left[
\left( \omega -2\right) \left( \omega -3\right) +12\alpha _{s}^{2}\right] }{%
\left( \omega -1\right) \left( \omega -2\right) \left( \omega
-3\right) +2\alpha _{s}^{2}\left( 7\omega -9\right) }$. Then one
can set the determinant $\left( \omega -1\right) \left( \omega
-2\right) \left( \omega -3\right) +2\alpha _{s}^{2}\left( 7\omega
-9\right) =0$ which yields three solutions. Each of those three
solutions stands for three poles $\omega _{1}=1+2\alpha _{s}^{2}$,
$\omega _{2}=2+10\alpha _{s}^{2}$ and $\omega _{3}=3-12\alpha
_{s}^{2}$, respectively. The inverse Laplace transform changes the
integer exponents to non-integer renormalized exponents,
\begin{eqnarray}
n_{2}^{(1)}(\alpha Y)=\frac{1+3\alpha _{s}^{2}}{1+\alpha _{s}^{2}}Ne^{\left(
1+2\alpha _{s}^{2}\right) \alpha Y}-\frac{2\alpha _{s}^{2}+100\alpha _{s}^{4}%
}{1-14\alpha _{s}^{2}}Ne^{\left( 2+10\alpha _{s}^{2}\right) \alpha Y}+\frac{%
72\alpha _{s}^{4}}{1-31\alpha _{s}^{2}}Ne^{\left( 3-12\alpha _{s}^{2}\right)
\alpha Y}.
\end{eqnarray}
\item $k=4$ case: One obtains $n_{4\times 4}^{(1)}(\omega )=\frac{N\left[
\left( \omega -2\right) \left( \omega -3\right) \left( \omega
-4\right) +24\alpha _{s}^{2}\left( 2\omega -5\right) \right]
}{\left( \omega -1\right) \left( \omega -2\right) \left( \omega
-3\right) \left( \omega -4\right) +2\alpha _{s}^{2}\left(
72-91\omega +25\omega ^{2}\right) +72\alpha _{s}^{4}} $. The
denominator has 4 poles which are $\omega _{1}=1+2\alpha
_{s}^{2}+o\left( \alpha _{s}^{4}\right) $, $\omega _{2}=2+10\alpha
_{s}^{2}+o\left( \alpha _{s}^{4}\right) $, $\omega _{3}=3+24\alpha
_{s}^{2}+1164\alpha _{s}^{4}$ and $\omega _{4}=4-36\alpha
_{s}^{2}-1080\alpha _{s}^{4}$. Then, the inverse Laplace transform
gives
\begin{eqnarray}
n_{3}^{(1)}(\alpha Y) &=&N\left( 1+o\left( \alpha _{s}^{2}\right) \right)
e^{\left( 1+2\alpha _{s}^{2}\right) \alpha Y}-2!N\alpha _{s}^{2}\left(
1+o\left( \alpha _{s}^{2}\right) \right) e^{\left( 2+10\alpha
_{s}^{2}\right) \alpha Y}  \notag \\
&&+3!N\alpha _{s}^{4}\left( 1+o\left( \alpha _{s}^{2}\right) \right)
Ne^{\left( 3+24\alpha _{s}^{2}\right) \alpha Y}-7344\left( \alpha
_{s}^{6}\right) Ne^{\left( 4-36\alpha _{s}^{2}\right) \alpha Y}.
\end{eqnarray}

\end{itemize}
\end{widetext}
From the above calculation for $n^{(1)}(\alpha Y)$, one can
already see that for an arbitrary $k$, the first $k-1$ poles are
stabilized to the
corresponding solutions $\omega _{j}=j+(3j^{2}-j)\alpha _{s}^{2}+\mathcal{O}%
\left( \alpha _{s}^{4}\right) $ $\left( 1\leq j\leq k\right) $ with a fixed
coefficient $j!N\left( \alpha _{s}^{2}\right) ^{j-1}$ and the last pole is
always $\omega _{k}=k-k\left( k-1\right) ^{2}\alpha _{s}^{2}+\mathcal{O}%
\left( \alpha _{s}^{4}\right) $ with a wrong coefficient since it misses the
recombination from $\left( k+1\right) $ particle mode as a result of
truncation at the $k$ particle mode. In addition, it is easy to see that the
$(3j^{2}-j)\alpha _{s}^{2}$ term is corresponding to the sum of the NLO
contributions $\left( j+1\right) j^{2}\alpha _{s}^{2}\alpha Y$ of the
leading loop diagrams (e.g. Fig.~\ref{reggeon}~B) and the leading
contributions ($-j\left( j-1\right) ^{2}\alpha _{s}^{2}\alpha Y$)\ of the
NLO loop diagrams (Fig.~\ref{reggeon}~C). On the other hand, the $-k\left(
k-1\right) ^{2}\alpha _{s}^{2}$ term in the last pole only corresponds to
the leading order contributions of the NLO loop diagrams (e.g. Fig.~\ref%
{reggeon}~C) as a consequence of the truncation at dimension $k$ which rules
out the leading loop diagram but allows the NLO loop diagrams at the $k$
particle mode.

\begin{widetext}
The pattern of the exact solution is transparent from the above discussion
and the generalization to infinite dimensions follows by simply taking the
limit $k\rightarrow \infty $. The exact solution for the $S$-matrix in
infinite dimensions reads
\begin{eqnarray}
S_{exact}(\alpha Y)=\dsum_{k=0}^{\infty }\left( -1\right)
^{k}\left( \alpha _{s}^{2}\right) ^{k}k!e^{\left( k+k\left(
3k-1\right) \alpha _{s}^{2}\right) \alpha Y+\mathcal{O}\left(
\alpha _{s}^{4}\right) \alpha Y}\left( 1+\mathcal{O}\left( \alpha
_{s}^{2}\right) \right) \ . \label{S_exact_2}
\end{eqnarray}
This result agrees with what we have found in the previous section, see Eq.~(%
\ref{smatrix}), however, it goes a step further since the $k\left(
3k-1\right) \alpha _{s}^{2}\alpha Y$ term is now in the exponent which can
be understood as a resummation of all $\alpha_S^2 \alpha Y$ terms via
exponentiation in Eq.~(\ref{smatrix}).
\end{widetext}

We notice that the $S$-matrix is not Borel-summable since $e^{3\alpha
_{s}^{2}\alpha Yk^{2}}$ is beyond the Borel-summability according to
Nevanlinna's theorem~\cite{hardy}. However, one can generalize the
Borel-summation technique and show that the above series still has a finite
sum by using analytic continuation. Generally speaking, one can see that the
$S$-matrix is the analytic function of two independent variables, $\alpha
_{s}^{2} $ and $\alpha Y$, when $\alpha _{s}^{2}$ is negative in the $\alpha
_{s}^{2}$ complex plane. Since $\alpha _{s}^{2}=0$ is not a singularity, one
can use analytic continuation to define the other half plane when $\alpha
_{s}^{2}$ becomes positive. In the following, we provide a generalized
Borel-summation method which yields a finite result.

\begin{figure}[htbp]
\begin{center}
\includegraphics[width=\columnwidth]{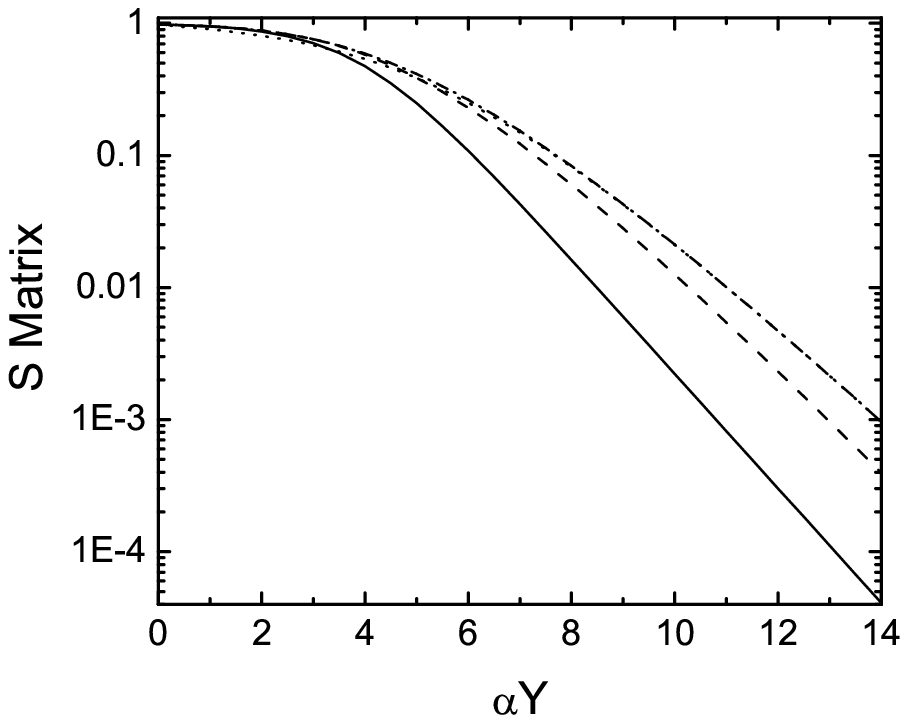}
\end{center}
\caption[*]{The $S$-matrix as a function of $\protect\alpha Y$ for $\protect%
\alpha _{s}^{2}=0.02$ in logarithmic scale: The solid line represents the
solution of the zero-transverse dimensional Kovchegov equation. The dashed
line corresponds to the leading order result, the dotted line stands for the
saddle point approximation ~(\protect\ref{exact2}) and the dot-dashed line
is extracted from the numerical evaluation of Eq.~(\protect\ref{num}).}
\label{tab_2}
\end{figure}

\begin{figure}[thbp]
\begin{center}
\includegraphics[width=\columnwidth]{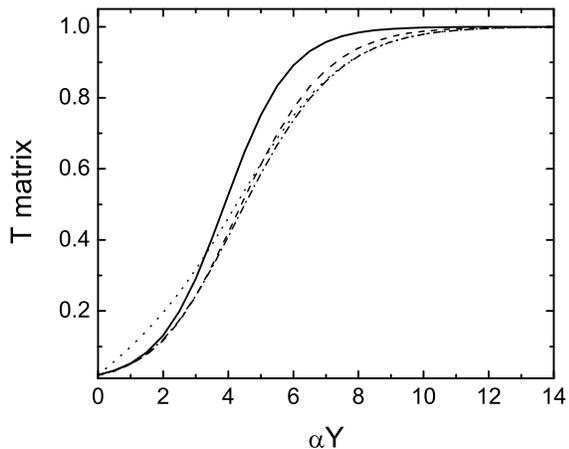}
\end{center}
\caption[*]{The $T$-matrix as a function of $\protect\alpha Y$ for $\protect%
\alpha _{s}^{2}=0.02$: The solid line represents the solution of the
zero-transverse dimensional Kovchegov equation. The dashed line corresponds
to the leading order result, the dotted line stands for the saddle point
approximation ~(\protect\ref{exact2}) and the dot-dashed line is extracted
from the numerical evaluation of Eq.~(\protect\ref{num}).}
\label{tmatrix}
\end{figure}

\begin{widetext}
We use a technique together with a subsequent Borel-summation in
order to maneuver the divergent series Eq.~(\ref{S_exact_2}) into
a definite and well-defined sum. Let us define $u=\alpha
_{s}^{2}e^{\left( 1-\alpha _{s}^{2}\right) \alpha Y}$ and $\gamma
=3\alpha _{s}^{2}\alpha Y$, then we obtain
\begin{eqnarray}
S(\alpha Y) &=&\dsum_{k=0}^{\infty }\left( -1\right)
^{k}k!u^{k}e^{\gamma k^{2}}, \\
&=&\dsum_{k=0}^{\infty }\left( -1\right) ^{k}\left( \frac{1}{u}%
\int_{0}^{\infty }dbb^{k}e^{-\frac{b}{u}}\right) \left( \sqrt{%
\frac{\gamma }{\pi }}\int_{-\infty }^{+\infty }dx\exp \left(
-\gamma
x^{2}+2\gamma xk\right) \right) , \\
&=&\sqrt{\frac{\gamma }{\pi }}\int_{-\infty }^{+\infty }dx\exp
\left( -\gamma x^{2}\right) \frac{1}{u}\int_{0}^{\infty }db\exp
\left( -\frac{b}{u}\right) \frac{1}{1+be^{+2\gamma x}},  \label{num} \\
&=&\sqrt{\frac{\gamma }{\pi }}\int_{-\infty }^{+\infty }dx\exp
\left( -\gamma x^{2}\right) \frac{1}{ue^{+2\gamma x}}\exp \left( \frac{1}{%
ue^{+2\gamma x}}\right) \Gamma \left( 0,\frac{1}{ue^{+2\gamma
x}}\right) \ .
\end{eqnarray}
To check the convergence of this integral, one can expand $\Gamma
\left( 0,z\right) $ at $z\rightarrow \infty $ which gives $\Gamma
\left( 0,z\right) \simeq \frac{\exp \left( -z\right) }{z}\left(
1-\frac{1}{z}\right) $, showing that the above integral is
definitely finite and well-defined. The $x$-integration can not be
performed analytically. However, one can use the saddle point
approximation to evaluate the above $x$-integral, in which one
finds the saddle point $x=-1$, and
\begin{eqnarray}
S(\alpha Y) &=&\sqrt{\frac{\gamma }{\pi }}\int_{-\infty }^{+\infty
}dx\exp \left( -\gamma x^{2}-2\gamma x\right) \frac{1}{u}\exp \left( \frac{1}{%
ue^{+2\gamma x}}\right) \Gamma \left( 0,\frac{1}{ue^{+2\gamma x}}\right) \\
&\approx &\frac{1}{\alpha _{s}^{2}e^{\left( 1-4\alpha _{s}^{2}\right) \alpha
Y}}\exp \left( \frac{1}{\alpha _{s}^{2}e^{\left( 1-7\alpha _{s}^{2}\right)
\alpha Y}}\right) \Gamma \left( 0,\frac{1}{\alpha _{s}^{2}e^{\left(
1-7\alpha _{s}^{2}\right) \alpha Y}}\right) .  \label{exact2}
\end{eqnarray}
\end{widetext}
A good agreement of this result with the numerical evaluation of Eq.~(\ref%
{num}) is shown in Fig.~(\ref{tab_2},\ref{tmatrix}).

Eq.~(\ref{exact2}) is a fairly exact result in the high-energy limit since
it presumably resums all the sub-leading diagrams up to $\alpha_{s}^{2}%
\alpha Y$ level, only neglects $\mathcal{O}\left( \alpha _{s}^{2}\right) $
(in the prefactor) and $\mathcal{O}\left( \alpha _{s}^{4} \alpha Y\right)$
(in the exponent) corrections. If we compare this with Eq.~(\ref{approx}) of
the previous section, we can see that Eq.~(\ref{approx}) is an approximate
solution which results from Eq.~(\ref{exact2}) when $\alpha _{s}^{2}\alpha
Y\ll 1$. The $\omega$-representation is equivalent to the integral
representation method in Sec.~\ref{integral_representation}, however, it
enables us to go one step further by resumming all the $\alpha
_{s}^{2}\alpha Y$ terms in the $S$-matrix. Thus, the $S$-matrix in Eq.~(\ref%
{exact2}) is assumed to be valid up to the rapidity $Y_{c2}=\frac{1}{\alpha
_{s}^{4}\alpha }$ where the next-to-next-to-leading order $\left( \mathcal{O}%
\left( \alpha _{s}^{4} \alpha Y\right)\right) $ becomes of order one.

\subsection{The Consequences of the NLO Correction}

In Mueller's toy model~\cite{Mueller:1994gb}, which was inspired by the QCD
dipole model (DM)~\cite{Mueller:1993rr+X}, the $S$-matrix for
particle-particle scattering in the center-of-mass frame is given by
%\begin{widetext}
\begin{eqnarray}
S^{DM}(Y) =\dsum_{m,n=1}^{\infty }e^{-\alpha_{s}^{2}mn}P_{m}\left(
\frac{Y}{2}\right) P_{n}\left( \frac{Y}{2}\right) \ ,
\label{borels}
\end{eqnarray}
%\end{widetext}
where $P_n(Y)$ is the probability density for having $n$ particles
in the wavefunction of an initial single particle after the
evolution up to $Y$. For large rapidities, $P_n(Y)$
reads~\cite{Mueller:1994gb}
\begin{eqnarray}
P(n,Y) =  \frac{1}{\overline{n}(Y)}\ e^{-n/\overline{n}(Y)}
\label{dist_ns}
\end{eqnarray}
with $\overline{n}(Y) = \exp(\alpha_s Y)$. Using $u =
n/\overline{n}(Y/2)$ and $v = m/\overline{n}(Y/2)$ and changing
the summations in eq.~(\ref{borels}) into integrals (this change
is rigorous since $du$=$dv$=$1/\overline{n}(Y/2)$ is
infinitesimal) for large rapidities, one obtains
\begin{eqnarray}
S^{DM}(Y) =\int_{\frac{1}{\overline{n}(Y/2)}}^{\infty }du
\int_{\frac{1}{\overline{n}(Y/2)}}^{\infty }dv\  \exp \left[
-\alpha _{s}^{2}\overline{n}\left( Y\right) uv-u-v\right] \
\label{dist_ns_int}
\end{eqnarray}
which after the integration over $u$ and $v$ leads to
\begin{eqnarray}
S^{DM}(Y) \simeq \frac{1}{\alpha _{s}^{2}e^{\alpha Y}}\ln \left[
\frac{\alpha
_{s}^{2}e^{\alpha Y}}{\left( 1+\alpha _{s}^{2}e^{\alpha Y/2}\right) ^{2}}%
\right] \ . \label{DM_result}
\end{eqnarray}
This result reduces for $Y \ll Y_s = \frac{2}{\alpha_s} \ln
\left(\frac{1}{\alpha_s^2}\right)$ to
\begin{eqnarray}
S^{DM}(Y) \simeq \frac{1}{\alpha _{s}^{2}e^{\alpha Y}}\ln \left[
\alpha _{s}^{2}e^{\alpha Y}\right] \label{Y_sY0}
\end{eqnarray}
while for $Y \gg Y_s$ it gets
\begin{eqnarray}
S^{DM}(Y) \simeq \frac{1}{\alpha _{s}^{2}e^{\alpha Y}}\ln \left[
  \frac{1}{\alpha_s^2} \right] \ .
\label{Y_lY0}
\end{eqnarray}
It is easy to see that the result in Eq.~(\ref{DM_result}) agrees with our
LO contribution given in Eq.~(\ref{approx}) only when $Y\leq Y_{s}$. For $%
Y\gg Y_{s}$ the result from Mueller's toy model given in Eq.~(\ref{Y_lY0})
can not be trusted since it does not include saturation effects within the
interacting particle wave functions which start becoming important at $%
Y\simeq Y_{s}$. Mueller's toy model fails to reproduce the LO contribution
of the LO diagrams (e.g. Fig~\ref{reggeon}~B) at $Y\gg Y_{s}$, it does not
include the NLO contribution of the LO diagrams (terms including $%
(k-1)^{2}\alpha Y/e^{\alpha Y}$ in Eq.~(\ref{LO_Loop})) and it also misses
the NLO loop diagrams (e.g. Fig~\ref{reggeon}~C).

Mueller and Salam~\cite{Mueller:1996te} have been able to
calculate the probability distribution $P_n(Y)$ which includes
saturation effects by imposing boost invariance on the small-x
evolution. They calculated the boost invariant S matrix in the
rest frame of the target, and obtained the same LO result in
Eq.~(\ref{approx}). We also find that it is interesting to
calculate the S matrix in the center of mass frame. Inserting
their result,
\begin{eqnarray}
P^{s}(n,y) = \exp \left[ \alpha_s^2\,n -\alpha\,y
-\frac{1}{\alpha_s^2}\,\left(e^{\alpha_s^2\,n}-1\right)e^{-\alpha
y}\right], \label{dist_ws}
\end{eqnarray}
in Eq.~(\ref{borels}), converting the summations into integrals, changing
the integration variables from $m-1$ to $m$ and from $n-1$ to $n$, setting $%
e^{\alpha_s^2} =1$, and using the integration variables $v = n/\overline{n}%
(Y/2)$ and $u = m/\overline{n}(Y/2)$, one ends up with the
boost-invariant result for the $S$-matrix in the center of mass
frame,
\begin{eqnarray}
S^{B}(Y)&\simeq&\int_{0}^{\infty }du \int_{0}^{\infty }dv\ \exp
\left[ -\alpha _{s}^{2}\overline{n}\left( Y\right) uv-u-v\right]
\label{limit_0} \\
&=& \frac{1}{\alpha_s^2 e^{\alpha\,Y}}\ e^{\frac{1}{\alpha_s^2 e^{\alpha\,Y}}%
} \ \Gamma(0,\frac{1}{\alpha_s^2 e^{\alpha\,Y}}) \label{res_ws}\\
&\simeq&
 \frac{1}{\alpha _{s}^{2}e^{\alpha Y}}\ln \left[ \alpha
_{s}^{2}e^{\alpha Y}\right] .  \label{res_ws1}
\end{eqnarray}
This result agrees with our LO contribution in Eq.~(\ref{approx}).
Note, however, that the Mueller-Salam toy model does not
incorporate the NLO correction given in Eq.~(\ref{approx}). Our
toy model, Eq.~(\ref{approx}), tells us that the NLO correction is
negligible for $Y \leq Y_c = \frac{1}{ \alpha\alpha_s^2}$. This
means that the result in Eq.~(\ref{res_ws}) is valid over a much
larger rapidity range, up to $Y \simeq Y_c = \frac{1}{
\alpha\alpha_s^2}$, as compared to the result in (\ref{DM_result})
which is
inspired from the dipole model and which breaks down at $Y_s=\frac{2}{%
\alpha_s} \ln \left(\frac{1}{\alpha_s^2}\right)$. Therefore, the
effect of saturation is to replace the result in Eq.~(\ref{Y_lY0})
by the result in Eq.~(\ref{Y_sY0}) even when $Y \geq Y_s$. It is
interesting to note that the saturation effects are mathematically
included by setting the lower limits
in Eq.~(\ref{dist_ns_int}) to $0$ as can be seen by comparing Eq.~(\ref%
{dist_ns_int}) with Eq.~(\ref{limit_0}) without modifying the
probability distribution.(In addition, one can easily prove that
the $S$ matrix of dipole approach is boost invariant if and only
if the lower limits in Eq.~(\ref{dist_ns_int}) are set to $0$.)

\section{Conclusion}

We have investigated a stochastic toy model without transverse dimensions
(equivalently, a infinite dimension hierarchy of evolution equation) which
naturally generates Pomeron loops. We have computed the Pomeron loop
diagrams to the NLO using two different methods. We find that the LO
calculation agrees with the exisiting results in the literature~\cite%
{Mueller:1994gb,Mueller:1996te}. The study of NLO graphs in this toy model
generates the $\left( \alpha _{s}^{2}\alpha Y\right) $ corrections.

Observations made by investigating Mueller's toy
model~\cite{Mueller:1994gb} have been shown numerically by
Salam~\cite{Salam:1995uy,Salam:1995zd} to be valid also in the
four-dimensional QCD. Presumably our toy model reveals properties
of the four-dimensional QCD evolution to some extent. We believe
that the NLO correction in the four-dimensional QCD would be of
order $\alpha _{s}^{2}\alpha Y$ as well (here we define $\alpha
=\alpha _{P}-1=\frac{ 4\alpha _{s}N_{c}}{\pi }\ln 2$ to be the
BFKL pomeron intercept). The behavior of our toy model is
heuristically indicating that NLO contributions can be neglected
in the four-dimensional QCD evolution as long as $Y \ll Y_c$ as
well. This believe is strengthened even more by noticing that the
NLO correction $\alpha^2_s \alpha Y$ seem to naturally appear in
QCD: There are two types of NLO contributions.

\begin{itemize}
\item The first type of NLO contribution comes from NLO
contribution of LO graphs(see example Fig.~\ref{reggeon}~B). For
the one-loop diagram, for instance, the LO contribution in QCD is
$\left( \alpha _{s}^{2}\right) ^{2}\exp \left( 2\alpha Y\right) $
which comes from varying the size of the loop from $0$ to $Y$. The
NLO contribution $\left( \alpha _{s}^{2}\right) ^{2}\alpha Y\exp
\left( \alpha Y\right) $ comes from changing the location of a
fixed zero-size loop from $0 $ to $Y$. The NLO contribution is
suppressed by the factor $\alpha Y/e^{\alpha Y}$ with respect to
the LO contribution, as in the toy model (see
Eq.~(\ref{LO_Loop})), leading to the general form $\alpha
_{s}^{2}\alpha Y$.

\item The second type of NLO contribution comes from LO contribution
of NLO graphs(see example Fig.~\ref{reggeon}~C). For the two-loop
diagram Fig.~\ref{reggeon}~C, for instance, the LO contribution
$\left( \alpha _{s}^{2}\right) ^{3}\alpha Y\exp \left(2\alpha
Y\right) $ comes from shifting the location of the fixed zero-size
connecting Pomeron between two loops from $0 $ to $Y$ while the
sizes of these two loops are changed in the same time. Since the
total amplitude of this graph does not depend on the location of
the connecting Pomeron, one can obtain an extra factor of $\alpha
Y$ by integrating from $0 $ to $Y$. The NLO contribution (see
Eq.~(\ref{NLOloop} )) is suppressed by the factor $\alpha
Y/e^{\alpha Y}$ with respect to the LO contribution of
Fig.~\ref{reggeon}~B, as in the toy model (see Eq.~(\ref{LO_Loop}
)), leading to the general form $\alpha _{s}^{2}\alpha Y$.
\end{itemize}

Therefore, we can see that the NLO correction generally takes the
form of $\alpha _{s}^{2}\alpha Y$. A QCD model in which all
Pomeron splittings occur before all Pomeron mergings in a
boost-invariant manner (as in this toy model) would lead to the
right LO result which would be valid up to the limit
$\frac{1}{\alpha \alpha _{s}^{2}}$.

\begin{acknowledgments}
We would like to thank Alfred Mueller for numerous stimulating and
insightful discussions. We are also grateful to Stephane Munier for many
useful remarks and comments. A. Sh. acknowledges financial support by the
Deutsche Forschungsgemeinschaft under contract Sh 92/2-1.
\end{acknowledgments}

\appendix

\section{The equivalence of splitting and recombination}

\label{duality}
\begin{figure}[tbp]
\begin{center}
\includegraphics[width=\columnwidth]{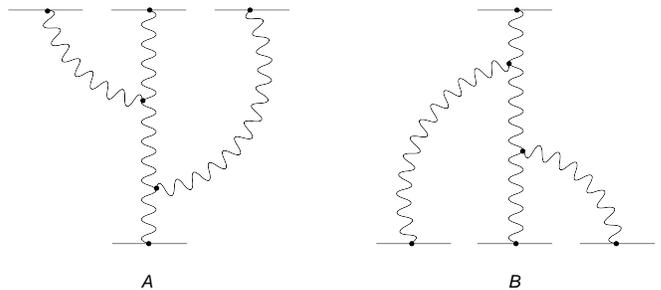}
\end{center}
\caption[*]{Graphs for one object scattering with three objects. Curly lines
represent BFKL ladders. In graph (A) a single particle evolves through
Pomeron splittings and scatters off three particles, while in graph (B)
three particles evolve through Pomeron mergings and scatter off a single
particle.}
\label{splitmerge}
\end{figure}
In this part, we show that Eq.~(\ref{h1}) gives Lorentz invariant results.
To show this, let us start with a single particle, then split it to three
particles which yields $n_{0}^{(3)}(y)=N3!e^{3\alpha y}(1-e^{-\alpha y})^{2}$
according to Eq.~(\ref{split}). Following Kovchegov~\cite{Kovchegov:1999yj},
the scattering amplitude is
\begin{eqnarray}
A(\alpha Y)=\dsum_{k=1}^{\infty }\frac{\left( -1\right) ^{k+1}}{k!}%
n_{0}^{(k)}(Y)T^{\left( k\right) }(0).  \label{amplitude}
\end{eqnarray}%
in which $T^{\left( k\right) }(0)=$ $\left( \alpha _{s}^{2}\right) ^{k}$ and
the $\frac{1}{k!}$ can be understood as the fact that those $k$-particles in
the target are the same. Therefore, it is very easy to see that $%
A_{s}(\alpha Y)=N\left( \alpha _{s}^{2}e^{\alpha y}\right) ^{3}(1-e^{-\alpha y})^{2}$
which corresponds to the diagram shown in Fig.~\ref%
{splitmerge}~A.

On the other hand, we can put the evolution into the target, let them merge
into one object before scattering with the projectile (Fig.~\ref{splitmerge}%
~B). This means that one starts with an initial condition $\left.
n_{0}^{(3)}\right\vert _{y=0}=N_{3}$ and $\left. n_{0}^{(k)}\right\vert
_{y=0}^{k\neq 3}=0$ in solving the Eq.~(\ref{h1}). After merging twice from $%
n_{0}^{(3)}(y)=N_{3}e^{3\alpha y}$, one obtains $\Delta
n^{(1)}=N_{3}\left( \frac{\beta }{\alpha }\right) ^{2}e^{3\alpha
y}(1-e^{-\alpha y})^{2}$. Finally, it is straight forward to see
that $A_{m}(\alpha Y)=N_{3}\left( \alpha _{s}^{2}e^{\alpha
y}\right) ^{3}(1-e^{-\alpha y})^{2}$
after setting $\frac{\beta }{\alpha }=\alpha _{s}^{2}$. Presumably, $%
A_{s}(\alpha Y)$ should be same as $A_{m}(\alpha Y)$ since they
actually describe the same process, and they are indeed the same
once we set $N=N_{3}$.

Moreover, it is tempting to replace the target by a large nucleus with $A$
nucleons inside it. Then, we need to define $T^{\left( k\right) }(0)=$ $%
\left( A^{\frac{1}{3}}\alpha _{s}^{2}\right) ^{k}$, and indeed one can see
that $A_{s}(\alpha Y)=A_{m}(\alpha Y)$ as long as we set $N_{3}=N\left( A^{%
\frac{1}{3}}\right) ^{3}$.
\begin{widetext}
\section{Solution to a slightly different stochastic differential equation}

\label{other_Langevin_equation} We notice that there has been a
lot of theoretical discussion and numerical simulation on a
slightly different stochastic differential equation which
has a noise term $\sqrt{2\left( \alpha \widetilde{n}-\beta \widetilde{n}%
^{2}\right) }\nu (y)$, namely the stochastic
Fisher-Kolmogorov-Petrovsky-Piscounov(sFKPP) equation. For
completeness, we provide the solution to this equation by using
the $\omega-$representation we developed above. The dynamics of
this equation is given by two random processes: by an increase
$dY$ in rapidity, a
particle can split into two particles with some rate $\alpha $ ($A\overset{%
\alpha }{\longrightarrow }A+A$) or two particles can merge into
one with a rate $2\beta $ ($A+A\overset{2\beta }{\longrightarrow
}A$). From the equation,
\begin{eqnarray}
\frac{d\widetilde{n}}{dy}=\alpha \widetilde{n}-\beta \widetilde{n}^{2}+\sqrt{%
2\left( \alpha \widetilde{n}-\beta \widetilde{n}^{2}\right) }\nu (y),
\label{Langevin2}
\end{eqnarray}%
one can get%
\begin{eqnarray}
\frac{dn^{(k)}}{dy}=\left( k\alpha -k\left( k-1\right) \beta
\right) n^{(k)}+k(k-1)\alpha n^{(k-1)}-k\beta n^{(k+1)}.
\label{hierarchy2}
\end{eqnarray}%
Similarly, one can write Eq.~(\ref{hierarchy2}) into a matrix in the $\omega
$-representation, then get the similar $S$-matrix of the scattering between
two dipoles,

\begin{eqnarray}
S(\alpha Y) &=&\dsum_{k=0}^{\infty }\left( -1\right) ^{k}\left(
\alpha _{s}^{2}\right) ^{k}k!e^{\left( k+2k^{2}\alpha
_{s}^{2}\right) \alpha Y+o\left( \alpha _{s}^{4}\right) \alpha
Y}\left( 1+o\left( \alpha
_{s}^{2}\right) \right) , \\
&=&\sqrt{\frac{\delta }{\pi }}\int_{-\infty }^{+\infty }dx\exp
\left( -\delta x^{2}\right) \frac{1}{ve^{+2\delta x}}\exp \left( \frac{1}{%
ve^{+2\delta x}}\right) \Gamma \left( 0,\frac{1}{ve^{+2\delta x}}\right)  \\
 &\approx &\frac{1}{\alpha _{s}^{2}e^{\left( 1-2\alpha
_{s}^{2}\right) \alpha Y}}\exp \left( \frac{1}{\alpha
_{s}^{2}e^{\left( 1-4\alpha _{s}^{2}\right) \alpha Y}}\right)
\Gamma \left( 0,\frac{1}{\alpha _{s}^{2}e^{\left( 1-4\alpha
_{s}^{2}\right) \alpha Y}}\right),
\end{eqnarray}
where $\delta=2\alpha _{s}^{2}\alpha Y$ and $v=\alpha
_{s}^{2}e^{\alpha Y}$. The difference of between this result and
the one in Eq.~(\ref{exact2}) is at the NLO correction, however,
one can see that Eq.~(\ref{Langevin}) has a more transparent
physical meaning in QCD since it contains the projectile-target
duality and one can easily relate it to the zero transverse
dimension reggeon theory.
\end{widetext}

\end{document}